\documentclass[oldversion,letter]{aa}
\usepackage{graphicx}
\usepackage{txfonts}
\usepackage{graphicx}
\usepackage{wasysym}
\begin{document}
   \title{Power for dry BL Lacertae objects}

   \author{A. Paggi
          \inst{1}
          \and
          A. Cavaliere\inst{1}
          \and
          V. Vittorini\inst{2}
          \and
          M. Tavani\inst{1}\({}^{,}\)\inst{2}
          }

   \institute{Dipartimento di Fisica, Universit\`{a} di Roma ``Tor Vergata", and\\
              INFN Roma Tor Vergata\\
              Via della Ricerca Scientifica 1, I-00133 Roma, Italy\\
              \email{paggi@roma2.infn.it}
         \and
             INAF/IASF-Roma, Via Fosso del Cavaliere 1, I-00100, Roma, Italy
             }

   \date{}


  \abstract
   {Is it significant that the intrinsic outputs of several BL Lacs are observed to level off at values of about \({10}^{46}\mbox{ erg}\mbox{ s}^{-1}\)? In searching for an answer, we compare \(\gamma\)-ray observations by the \emph{AGILE} satellite of the BL Lac S5 0716+714 with those of Mrk 421 and Mrk 501; the former are particularly marked by intense flares up to fluxes of \(2\times{10}^{-6}\mbox{ photons}\mbox{ cm}^{-2}\mbox{ s}^{-1}\) in the \(0.1-10\mbox{ GeV}\) energy range.
   These ``dry" BL Lacs show evidence of neither thermal disk emissions nor emission lines signaling any accreting or surrounding gas; the spectral distributions of their pure non-thermal radiations are effectively represented by the synchrotron self-Compton process.
   With source parameters correspondingly derived and tuned with simultaneous multiwavelength observations, we find for S5 0716+714 a total jet power of about \(3\times{10}^{45}\mbox{ erg}\mbox{ s}^{-1}\), which makes it one of the brightest dry BL Lacs so far detected in \(\gamma\) rays.
   We evaluate the mass of the associated Kerr hole to be around \(5\times{10}^8 M_{\tiny{\astrosun}}\), implying that the source is significantly gauged in terms of the
   maximal power around \(4\times{10}^{45}\mbox{ erg}\mbox{ s}^{-1}\) extractable via the Blandford-Znajek electrodynamical mechanism; other dry BL Lacs observed
    in \(\gamma\) rays remain well below that threshold. These findings and those forthcoming from \emph{Fermi}-LAT will provide a powerful test of electrodynamics
    in the surroundings of the hole, that are dominated by GR effects.}


   \maketitle
%

\section{Introduction}\label{sezione1}

   Blazars rank among the brightest active galactic nuclei on the basis of their inferred isotropic luminosities that may attain some \(L_{iso}\sim {10}^{48}\mbox{ erg}\mbox{ s}^{-1}\). Actually, these sources radiate from a narrow relativistic jet closely aligned with the observer's line of sight. The jet emits highly beamed non-thermal radiations, with observed fluxes enhanced by aberration and Doppler effects of Special Relativity (Begelman, Blandford \& Rees \cite{bbr}; K\"{o}nigl \cite{konigl86}; Urry \& Padovani \cite{urry}). So the luminosities \(L_{iso}\) greatly exceed the intrinsic outputs of the jets that easily level off at \({10}^{-2} - {10}^{-3} L_{iso}\).

   The BL Lac objects (henceforth BL Lacs) in particular are blazars that show no or just weak and intermittent emission lines. Their spectra are represented well as a continuous spectral energy distribution (SED) \(S_\nu=\nu F_\nu\) featuring two peaks: one at a lower frequency due to synchrotron emission by highly relativistic electrons; and a higher frequency counterpart due to inverse Compton upscattering by the same electron population of seed photons provided by the synchrotron emission itself (synchrotron self-Compton, SSC; see Jones, O'Dell \& Stein \cite{jones}, Marscher \& Gear \cite{marscher}; Maraschi, Ghisellini \& Celotti \cite{maraschi}), with possible additions from sources external to the beam (external Compton, EC; see Dermer \& Schlickeiser \cite{dermer93}; Sikora, Begelman \& Rees \cite{sikora}).

   The BL Lacs also exhibit strong variability on timescales of days to minutes with substantial flux variations (flares) particularly at high
   energies as realized early on (see Setti \& Woltjer \cite{setti}, and references therein).

   Here we focus on ``dry" BL Lacs, that is, sources with no evidence of surrounding gas, such as emission lines or a big blue bump (see Peterson \cite{peterson},
   Kembhavi \& Narlikar \cite{kembhavi}) related to current accretion. They provide an appropriate testing ground for comparing their intrinsic outputs with maximal powers extractable from rotating supermassive black holes and from the dragged accretion disks by means of large-scale electromagnetic fields, via the intriguing, variously debated Blandford-Znajek electrodynamics (BZ, Blandford \& Znajek \cite{bz}; see also Ghosh \& Abramowicz \cite{ghosh}; Krolik \cite{krolik}; Livio, Ogilvie \& Pringle \cite{livio}; Cavaliere \& D'Elia \cite{cavaliere}; McKinney \cite{mckinney05}; Nemmen et al. \cite{nemmen}; Tchekhovskoy, McKinney \& Narayan \cite{thcek}).
     The bare hole contribution can yield up to
     \(L_K \sim 2\times{10}^{45}\left({{M_{\medbullet}}/{{10}^9\,M_{\tiny{\astrosun}}}}\right)\mbox{ erg}\mbox{ s}^{-1}\),
     given the hole mass \(M_{\tiny{\medbullet}}\) in units of \({10}^9M_{\tiny{\astrosun}}\) and a magnetic field  \(B\sim{10}^4\mbox{ G}\) threading its horizon.

     In the following, we adopt the standard, flat cosmology with \(H_0 = 72 \mbox{ km}\mbox{ s}^{-1}\mbox{ Mpc}^{-1}\) and \(\Omega_\Lambda = 0.74\) (Dunkley et al. \cite{dunkley}).


\section{The radiation process}\label{radproc}

The SEDs of the dry BL Lacs are widely understood in terms of the simple, homogeneous SSC process. This is based on radiations produced in a region containing a magnetic field and relativistic electrons accelerated to high random energies \(\gamma m c^2\) (with \(\gamma\) up to \({10}^6 - {10}^8\)) that move toward the observer with bulk Lorentz factors \(\Gamma\sim 10 - 20\) (see Ghisellini et al. \cite{ghisellini}).

To begin with, we assume the sources to have an isotropic geometry with a radius \(R\) as a single size parameter, and to contain the relativistic electrons and non-relativistic protons with the same \(\Gamma\), at a common density \(n\). Observed (primed) frequencies and fluxes are related to the rest frame (unprimed) quantities by means of \(\nu'=\nu\,\delta\) and \(F'= F\, \delta^4\) (Begelman, Blandford \& Rees \cite{bbr}), where \(\delta={\left[{\Gamma\left({1-\beta\cos{\theta}}\right)}\right]}^{-1}\) is the beaming factor related to the angle \(\theta\) between the jet and the line of sight. Small viewing angles \(\theta\sim 1/\Gamma\) yield \(\delta\approx 2 \Gamma\). Correspondingly, the intrinsic variability and crossing times \(R/c\) will be longer than the observed ones \(R/\delta c\).

On empirical and theoretical grounds, we adopt log-parabolic shapes for the electron energy distributions. These are obtained from a Fokker-Planck equation in the presence of systematic and stochastic acceleration processes as first shown by Kardashev (\cite{kardashev}) and computed in detail by Paggi et al. (\cite{paggi}); the acceleration times scale as \(t_a\propto \gamma / E\), in terms of the effective electric field \(E\) (Cavaliere \& D'Elia \cite{cavaliere}). We therefore write the electron distribution in the form
\begin{equation}
N(\gamma)=N_0\,
{\left({\frac{\gamma}{\gamma_0}}\right)}^{-s-r\log{\left({\frac{\gamma}{\gamma_0}}\right)}}\, ,
\end{equation}
where \(s\) is the constant contribution to the slope, \(r\) is the ``curvature", and \(\gamma_0\) is the injection energy.
The energetic content of such an electron population can be expressed in terms of the rms adimensional energy
\(\gamma_p={\left({{\int{\gamma^2 N{\left({\gamma}\right)}\, d\gamma}}/{\int{N{\left({\gamma}\right)}\, d\gamma}}}\right)}^{1/2}\), to yield an energy density close to \(n\,\gamma_p\, m c^2\).

The emitted synchrotron SED is correspondingly given by (Massaro et al. \cite{massaro2}; Tramacere et al. \cite{tramacere}, and references therein)
\begin{equation}\label{logpar}
S_\nu=S_0
\,{\left({\frac{\nu}{\nu_0}}\right)}^{-a-b\log{\left({\frac{\nu}{\nu_0}}\right)}}\, ,
\end{equation}
with a constant contribution \(a=(s-3)/2\) to the spectral index, a spectral curvature \(b\approx r/5\), and a peak frequency \(\nu_S\propto B\gamma_p^2\). For the IC, an analogous SED applies, where in the Thomson regime one has \(a=(s-3)/2\), \(b\approx r/10\) and a peak frequency \(\nu_C\propto B\gamma_p^4\), whilst in the Klein-Nishina regime \(a=s\), \(b\approx r\), and \(\nu_C\propto \gamma_p\) hold.

\begin{figure}
\includegraphics[scale=0.33]{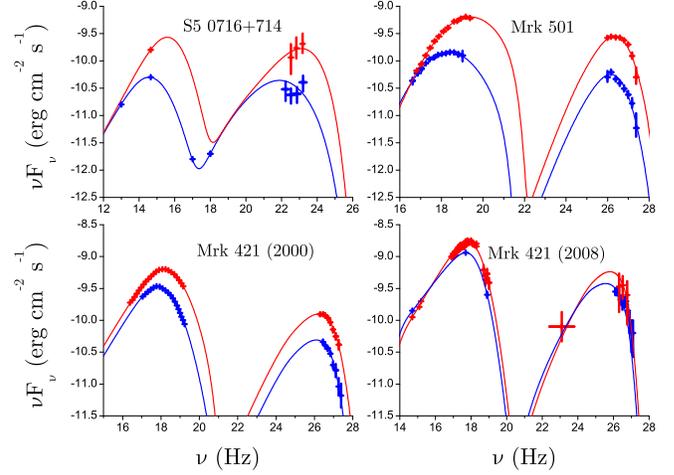}
  \caption{SEDs of the BL Lacs considered in the text: S5 0716+174 (upper-left frame), Mrk 501 (upper-right frame) and Mrk 421 (lower-left and lower-right frames), each in a low (blue line) and a high (red line) state. In terms of intrinsic luminosity S5 0716+714 is the strongest source (data referenced in Sect. \ref{radproc}).}\label{blazars}
\end{figure}

In these SEDs and distributions, the peaks where most of the energy resides are not materially affected by radiative cooling active on timescales \(t_c\propto 1/\gamma_p\) longer than the crossing time \(R/c\); cooling will rather erode the high energy tails.
As an added bonus, the synchrotron and IC radiations from a log-parabolic electron population irreversibly \emph{broaden} under the action of the stochastic acceleration component, following \(b\propto 1/t\) (Paggi et al. \cite{paggi}); thus, a sudden increase in the spectral curvature will mark the emergence of a new electron population.
Last but not least, Eq. \ref{logpar} closely fits (as we illustrate in Fig. \ref{blazars}) the spectra of the sources that we focus on in view of extended spectral coverage provided by their multiwavelength, \emph{simultaneous} observations.
%

This is the case for S5 0716+714, which has the following data available: a low state provided by historical AIT and EGRET data (Lin et al. \cite{lin});
a high state in September 2007, in the \(\gamma\)-ray range covered by \emph{AGILE}-GRID where the IC peaks for this ``intermediate" BL Lac (Padovani \& Giommi \cite{pg}); optical and radio data taken with GASP-WEBT on September 7-12 (Villata et al. \cite{villata}; Vittorini et al. \cite{vittorini}). The multiwavelength variations observed by Giommi et al. (\cite{giommi}) and the increased spectral curvature (see Fig. \ref{blazars}) in 2007 are indicative of the injection of a second component.

The source may be compared with Mrk 501, for which we consider two states with simultaneous BeppoSAX and CAT observations on April 7 and 16, 1997 (Massaro et al. \cite{massaro}); and with Mrk 421, for which we have low and high states in 2000 from BeppoSAX and HEGRA data (Konopelko et al. \cite{konopleko}), and multiwavelength observations performed with GASP-WEBT, \emph{RXTE}/ASM, \emph{Swift}, SuperAGILE, \emph{AGILE}-GRID, ARGO-YBJ, and VERITAS in June 2008 (Donnarumma et al. \cite{donnarumma}; Di Sciascio et al. \cite{disciascio}).

\section{The source power}

We are interested in the \emph{intrinsic} outputs referred to the jet frame, rather than in the luminosities \(L_{iso}= 4\pi\, D_L^2\, F\) inferred from insisting on an isotropic distribution of the observed flux \(F\), at the luminosity distance \(D_L\). We assume one ``cold" proton per electron satisfying \(\left\langle{\gamma}\right\rangle \la m_p / m_e\) (with the average \(\left\langle{\gamma}\right\rangle=\gamma_p\times {10}^{-1/4r}\) bounded in terms of the electron \(m_e\) and the proton \(m_p\) masses), and follow Celotti \& Ghisellini (\cite{celotti}) in writing for the intrinsic radiative luminosity \(L_r\) contributed by both the synchrotron and IC radiations and for the related powers carried by the jet, the expressions
\begin{eqnarray}
\label{lumprima}L_r &=& L_{iso}\,{\Gamma^2}/{\delta^4}\approx L_{iso}/16\,\Gamma^2\,,\\
L_e &=& \frac{4}{3}\pi\,R^2\,c\,n\,m_e\,c^2\left\langle{\gamma}\right\rangle\Gamma^2\,,\\
\label{lumultima}L_p &\sim& L_e\, {m_p}/{m_e\left\langle{\gamma}\right\rangle}\, ,\qquad L_B \ll L_{r,\,e,\,p}\,.
\end{eqnarray}
The total jet power is therefore given by \(L_{T}=L_r+L_e+L_p+L_B\); this is dominated by \(L_r\) and by \(L_e\), with \(L_B < L_p \la L_e\) for the fields \(B\sim 0.1 - 1\mbox{ G}\) implied by the spectral fits.

The simultaneous, multiwavelength observations enable extended spectral fits to determine the five key observables (beside the spectral curvature \(b\)) from the SSC model, namely: the synchrotron peak frequency and flux, the IC peak frequency and flux, and the variation time (see Paggi et al. \cite{paggi}). These lead to robust evaluations of the five source parameters \(n\), \(R\), \(B\), \(\left\langle{\gamma}\right\rangle\), and \(\delta\) (or \(\Gamma\)) entering Eqs. \ref{lumprima} - \ref{lumultima}; the main parameters are collected in Table \ref{tabella}.

\begin{table}
\centering
\begin{tabular}{|c|c|c|c|c|}
\hline
Source name & \(\delta\) &\(\gamma_p\) & \(L_{r}\) & \(L_{T}\)
\\
\hline
S5 0716+714 (low) & 12 & \(9.1\times{10}^{2}\) & \(4.2\times{10}^{44}\) & \(1.3\times{10}^{45}\)
\\
S5 0716+714 (high) & 19 & \(2.8\times{10}^{3}\) & \(1.5\times{10}^{45}\) & \(3.1\times{10}^{45}\)
\\
Mrk 501 (low)  & 15 & \(1.4\times{10}^{5}\) & \(4.9\times{10}^{42}\) & \(1.2\times{10}^{43}\)
\\
Mrk 501 (high) & 15 &\(1.9\times{10}^{5}\) & \(2.3\times{10}^{43}\) & \(3.1\times{10}^{43}\)
\\
Mrk 421 (2000, low)  & 20 &\(5.2\times{10}^{4}\) & \(4.0\times{10}^{42}\) & \(1.9\times{10}^{43}\)
\\
Mrk 421 (2000, high) & 20 &  \(6.1\times{10}^{4}\) & \(8.1\times{10}^{42}\) & \(3.1\times{10}^{43}\)
\\
Mrk 421 (2008, low)  & 20 & \(2.5\times{10}^{4}\) & \(1.6\times{10}^{43}\) & \(4.1\times{10}^{43}\)
\\
Mrk 421 (2008, high) & 20 & \(3.4\times{10}^{4}\) & \(2.1\times{10}^{43}\) & \(4.6\times{10}^{43}\)
\\
\hline
\end{tabular}
\caption{Parameters for the BL Lac sources discussed in the text. The redshifts are \(z=0.031\) for S5 0716+714, \(z=0.034\) for Mrk 501 and \(z=0.030\) for Mrk 412; \(L_r\) and \(L_T\) are given in \(\mbox{erg}\mbox{ s}^{-1}\).}
\label{tabella}
\end{table}

\section{The BZ benchmark}

As anticipated in Sect. \ref{sezione1}, a natural benchmark for these powers is provided by the BZ mechanism for electrodynamical energy extraction from a Kerr hole spun up to maximal rotation by past accretion episodes. A minimal, vestigial disk is required to hold the poloidal magnetic field threading the horizon; the disk is kept active by low accretion rates \(\dot{m}\la {10}^{-2}\) in Eddington units, loses angular momentum mainly via the large-scale field, and contributes some \(3\, L_K\) to the total power (Blandford \& Znajek \cite{bz}; Livio et al. \cite{livio}). The two contributions add to yield
\begin{equation}\label{BZ}
L_{BZ} \approx 8\times{10}^{45}\left({\frac{M_{\tiny{\medbullet}}}{{10}^9\,M_{\tiny{\astrosun}}}}\right)\mbox{ erg}\mbox{ s}^{-1}\,.
\end{equation}
We note that the balance \(B^2 /4\pi \sim p\) between the magnetic stress and the kinetic or radiation pressure \(p\) in the disk yields \(B\sim {10}^4\mbox{ G}\); for a radiation-pressure dominated disk, we have at the inner rim \(B^2\propto 1/M_{\tiny{\medbullet}}\), so \(B\) has dropped out of Eq. \ref{BZ}.

The hole mass is then the key parameter, that we evaluate from its correlation with the absolute red magnitude \(M_R\) of the host galactic bulge (Ferrarese et al. \cite{ferrarese}; Gebhardt et al. \cite{gebhardt}; Falomo et al. \cite{falomo2003}); for our cosmology this reads
\begin{equation}\label{falomo}
\log{\left({\frac{M_{\tiny{\medbullet}}}{M_{\tiny{\astrosun}}}}\right)} = -0.50\,M_R -2.61\, ,
\end{equation}
with scatter \(\pm 0.4\mbox{ dex}\) (Bettoni et al. \cite{bettoni}). For the host galaxy of S5 0716+714, observations of the magnitude \(R=18.3\pm 0.5\) reported by Nilsson et al. (\cite{nilsson}), besides indicating the redshift \(z=0.31\pm0.08\), yield a mass
\(M_{\tiny{\medbullet}}\simeq {5.5 ^{+8.0}_{-3.3}}\,{10}^8\,M_{\tiny{\astrosun}}\); the central value is consistent with estimates from microvariability of the optical flux (Sasada et al. \cite{sasada}),
which yield \(M_{\tiny{\medbullet}} \sim {10}^8 M_{\tiny{\astrosun}}\). For Mrk 501 and Mrk 421, one obtains \(M_{\tiny{\medbullet}}\simeq {1.0 ^{+2.4}_{-0.7}}\,{10}^9\,M_{\tiny{\astrosun}}\) and \(M_{\tiny{\medbullet}}\simeq {4.1 ^{+7.8}_{-2.7}}\,{10}^8\,M_{\tiny{\astrosun}}\), respectively.

\begin{figure}
\includegraphics[scale=0.35]{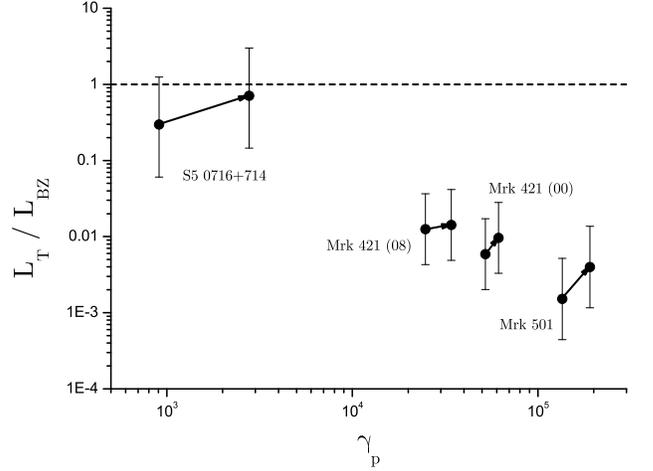}
  \caption{For the named sources the total jet luminosities normalized to their BZ power are plotted against the electron rms energy. Bars represent the hole mass uncertainties reflecting those (at the \(1-\sigma\) level) in host bulge luminosities, and including the scatter in Eq. (\ref{falomo}).}\label{fig1}
\end{figure}

Our results normalized to the respective \(L_{BZ}\) from Eq. \ref{BZ} are represented in Fig. \ref{fig1}. During flares, the electron rms energies (and the peak frequencies) are boosted in all sources, and so are the luminosities; this indicates that rising flares are directly related to increased \emph{acceleration} of the emitting electrons.

We emphasize that the powerful source S5 0716+714 is apparently constrained to move \emph{sideways}, as if to skirt the BZ limit; Mrk 421 in the 2008 states exhibits a similar behaviour, although with lower significance. On the other hand, the weaker source Mrk 421 in 2000 and the yet weaker Mrk 501 remain considerably below the BZ limit and so are expected to be free to move more vertically, as they do.

\section{Discussion and conclusions}

For dry BL Lacs with accretion rates \(\dot{m}<{10}^{-2}\), the SSC radiation process provides a robust evaluation of the jet luminosities. Whence we conclude that \(L_{BZ}\) provides a significant \emph{benchmark} for the output of the BL Lacs discussed here, and indeed  an upper limit to both their quiescent states and flares. In fact, during a recent flare S5 0716+714 was observed to be constrained by \(L_{BZ}\lesssim{10}^{46}\mbox{ erg}\mbox{ s}^{-1}\), and a similar behavior was observed in 2008 for Mrk 421. Non-thermal, beamed powers in the range \(L_K - L_{BZ}\) also provide evidence of an accretion disk that is active mainly in launching and channeling the jets by means of large-scale fields.

\begin{figure*}
\includegraphics[scale=0.40]{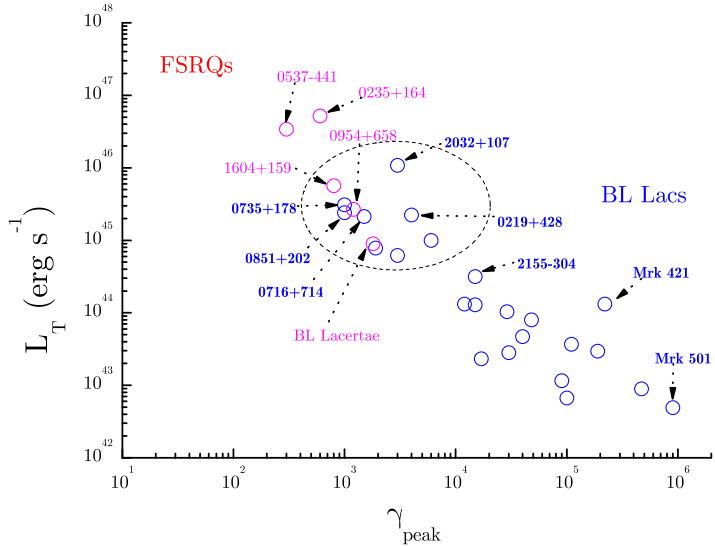}
  \caption{Bright BL Lacs in their context, adapted from Celotti \& Ghisellini (\cite{celotti}) with historical data in terms of \(L_T\) and the energy \(\gamma_{peak}\) related to the synchrotron peak (\(\gamma_{peak}=\gamma_{p}\times {10}^{1/2r}\)). Blue circles indicate dry, while violet circles indicate wet BL Lacs. The lower-left region of the diagram corresponds to the source condition \(t_c > R/c\), and the upper-right to \(t_c\gtrsim t_a\). Bright FSRQs lie at lower \(\gamma_{peak}\) and higher powers, with increasing signs of current \(\dot{m}\rightarrow 1\); selection effects depopulate weaker sources in the lower region (see Padovani \cite{padovani07}). The dashed oval highlights sources in the transition region from dry to wet BL Lacs, interesting to compare with \(L_{BZ}\) from Eq. \ref{BZ} particularly when at \(z\gtrsim 0.3\).}
\label{bs}
\end{figure*}

Referring to Fig. \ref{bs} and its caption, we note that during flares the sources move in the \(L_{T}\) - \(\gamma_{peak}\) plane away from the envelope that is outlined by bright BL Lacs with increasing rates \(\dot{m}\); the envelope ends up in the locus of the yet brighter Flat Spectrum Radio Quasars (FSRQs) with \(\dot{m}\sim 1\). The flares then move into a region of faster radiative cooling (Celotti \& Ghisellini \cite{celotti}, and references therein). This implies short-lived flares on timescales \(\la 1\mbox{ day}\), or requires shorter acceleration times \(t_a\sim \gamma /E\) with higher \(E\), as an alternative to structured sources such as decelerating (Georganopoulos \& Kazanas \cite{gk}) or spine-sheath jets with inner scale \(R_1<R\) (Tavecchio \& Ghiselllini \cite{tavecchio}). Faster acceleration and deviations from the envelope are consistent with flares caused by electron \emph{boost} rather than episodes of increased accretion onto the disk.

In this context, we recall (see Blandford \cite{blandford}; Padovani et al. \cite{padovani}) that sources lying along the envelope in Fig. \ref{bs} at higher \(L\) and lower \(\gamma_{peak}\) often exhibit stronger evidence of current accretion up to \(\dot{m}\sim 1\), such as thermal emissions and surrounding gas (big blue bump and broad emission lines), with a larger contribution from EC.
In fact, the progression from dry BL Lacs to FSRQs is likely to involve an enhanced and extended disk contribution as described by Blandford \& Payne (\cite{bp}), starting with ``wet" BL Lacs with \(\dot{m}\sim{10}^{-1}\); these feature larger EC contributions (Dermer et al. \cite{dermer}) and looming evidence of gas, including some thermal disk emission and weak or intermittent lines (Celotti, Ghisellini \& Fabian \cite{celotti07}). The last step in this progression is constituted by the powerful FSRQs with extant broad lines, a big blue bump from disks accreting at full rates \(\dot m \sim 1\), and a dominant or towering EC (Maraschi \& Tavecchio \cite{maraschi01}).

We add that the outputs of even misaligned BL Lacs may be \emph{calorimetrically} gauged from their feedback actions on the intra-cluster plasma surrounding their host galaxy when located in a cluster or a group of galaxies, as discussed by McNamara et al. (\cite{mcnamara}). These authors evaluate average powers around \({10}^{46}\mbox{ erg}\mbox{ s}^{-1}\) injected into the cluster MS0735.6+7421, and possibly also in the cluster A2029 and the group AWM 4.

The whole of the above evidence provides observational support to the \emph{relevance} of the electrodynamical BZ mechanism, and invites extended sampling of other interesting sources (see Fig. \ref{bs}).

If in dry BL Lacs with \(M_{\tiny{\medbullet}}< {10}^9 M_{\tiny{\astrosun}}\) the \(L_{BZ}\) limit were found to be substantially exceeded by outputs \(L_T> {10}^{46}\mbox{ erg}\mbox{ s}^{-1}\), this would require \(B > {10}^4\mbox{ G}\) at the Kerr horizon. These fields imply large dynamical stresses bounded only by \(B^2/4 \pi  \leq\rho c^2\), associated with particle orbits plunging from the disk toward the hole horizon (Meier \cite{meier2}) into a region fully controlled by strong gravity effects.

Thus, all such sources will provide powerful tests for the coupling of electrodynamics with General Relativity in full swing, and constitute an exciting arena for \emph{AGILE} and \emph{Fermi}-LAT data.

\begin{acknowledgements}
      We are grateful to R. Falomo for useful discussions concerning \(M_{\medbullet}\) evaluations for BL Lacs, and specifically in S5 0716+714.
      We acknowledge our referee for useful comments and helpful suggestions.
\end{acknowledgements}


\begin{thebibliography}{}

\bibitem[1984]{bbr}Begelman, M. C., Blandford, R. D. \& Rees, M. J. 1984, RvMP, 56, 255
\bibitem[2003]{bettoni}Bettoni, D., Falomo, R., Fasano, G., et al. 2003, A\&A, 399, 869
\bibitem[1990]{blandford}Blandford, R. D. 1990, in Saas-Fee Advanced Course 20, Active Galactic Nuclei, ed. R. D. Blandford, H. Netzer, \& L. Woltjer (Springer), 161
\bibitem[1982]{bp}Blandford, R.D., \& Payne, D. G. 1982, MNRAS, 199, 883
\bibitem[1977]{bz}Blandford, R. D., \& Znajek, R. L. 1977, MNRAS, 179, 433
\bibitem[2002]{cavaliere}Cavaliere, A., \& D'Elia, V. 2002, ApJ, 571, 226
\bibitem[2007]{celotti07}Celotti, A., Ghisellini, G. \& Fabian, A. C. 2007, MNRAS, 375, 417
\bibitem[2008]{celotti}Celotti, A. \& Ghisellini, G. 2008, MNRAS, 385, 283
\bibitem[2008]{dermer}Dermer, C. D., Finke, Justin, D. \& Menon, G. 2008 [arXiv:0810.1055]
\bibitem[1993]{dermer93}Dermer, C. D. \& Schlickeiser, R. 1993, ApJ, 416, 458
\bibitem[2009]{disciascio}Di Sciascio, G. and the ARGO-YBJ Collaboration 2009, Mem. S.A.It., 75, 282 [arXiv0907:2526]
\bibitem[2009]{donnarumma}Donnarumma, I., et al. 2009, ApJ, 691, L13
\bibitem[2009]{dunkley}Dunkley, J., Komatsu, E., Nolta, M. R., et al. 2009, ApJS, 180, 306
\bibitem[2003]{falomo2003}Falomo, R., et al. 2003, ApJ, 595, 624
\bibitem[2000]{ferrarese}Ferrarese, L., \& Merritt, D. 2000, ApJ, 539, L9
\bibitem[2000]{gebhardt}Gebhardt, K., et al. 2000, ApJ, 539, L13
\bibitem[2003]{gk}Georganopoulos, M., \& Kazanas, D. 2003, ApJ, 594, 27
\bibitem[1993]{ghisellini}Ghisellini, G., Padovani, P., Celotti, A., et al. 1993, ApJ, 407, 65
\bibitem[1997]{ghosh}Ghosh, P., \& Abramowicz, M. 1997, MNRAS, 292, 887
\bibitem[2008]{giommi}Giommi, P., et al. 2008, A\&A, 487, L49
\bibitem[1974]{jones}Jones, T. W., O'Dell, S. L., \& Stein, W. A. 1974, ApJ, 188, 353
\bibitem[1962]{kardashev}Kardashev, N. S. 1962, SvA, 6, 317
\bibitem[1999]{kembhavi}Kembhavi, A. K., Narlikar, J. V. 1999, \textit{Quasars and Active Galactic Nuclei}, Cambridge, Cambridge University Press
\bibitem[1986]{konigl86}K\"{o}nigl, A. 1986, NYASA, 470, 88
\bibitem[1999]{krolik}Krolik, J. H. 1999, ApJ, 515, 73
\bibitem[2003]{konopleko}Konopelko, A., et al 2003, ApJ, 597, 851
\bibitem[1995]{lin}Lin, Y.C., et al. 1995, ApJ, 442, 96
\bibitem[1999]{livio}Livio, M., Ogilvie, G., \& Pringle, J. 1999, ApJ, 512, 100
\bibitem[2009]{mcnamara}McNamara, B. R., Kazemzadeh, F., Rafferty, D. A., et al. 2007, ApJ, 698, 594
\bibitem[1992]{maraschi}Maraschi, L., Ghisellini, G., \& Celotti, A. 1992, ApJ, 397, L5
\bibitem[2001]{maraschi01}Maraschi, L., \& Tavecchio, F. 2001, ASPC, 227, 40M
\bibitem[1985]{marscher}Marscher, A. P., \& Gear, W. K. 1985, ApJ, 298, 114
\bibitem[2004]{massaro2}Massaro, E., et al. 2004, A\&A, 413, 489
\bibitem[2006]{massaro}Massaro, E., et al. 2006, A\&A, 448, 861
\bibitem[2004]{mckinney04}McKinney, J. C. \& Gammie, C. F. 2004, ApJ, 611, 977
\bibitem[2005]{mckinney05}McKinney, J. C. 2005, ApJ, 630, 5
\bibitem[1999]{meier}Meier, D. L. 1999, ApJ, 522, 753
\bibitem[2002]{meier2}Meier, D. L. 2002, NewAR, 46, 247
\bibitem[2007]{nemmen}Nemmen, R. S., Bower, R. G., Babul, A., et al. 2007, MNRAS, 377, 1652
\bibitem[2008]{nilsson}Nilsson, K., et al. 2008, A\&A, 487, L29
\bibitem[2009]{paggi}Paggi, A., et al. 2009, A\&A, 504, 821
\bibitem[1995]{pg}Padovani, P. \& Giommi, P. 1995, ApJ, 444, 567
\bibitem[2007]{padovani07}Padovani, P. 2007, Ap\&SS, 309, 63
\bibitem[2007]{padovani}Padovani, P., Giommi, P., Landt, H., et al. 2007, ApJ, 662, 182
\bibitem[1997]{peterson}Peterson, B. M. 1997, \textit{An Introduction to Active Galactic Nuclei}, Cambridge, Cambridge University Press
\bibitem[2008]{sasada}Sasada, M., et al. 2008, PASJ, 60, 37
\bibitem[1994]{sikora}Sikora, M., Begelman, M. C., \& Rees, M. J. 1994, ApJ, 421, 123
\bibitem[1994]{setti}Setti, G., Woltjer, L. 1994, ApJS, 92, 629
\bibitem[2008]{tavecchio}Tavecchio, F., \& Ghisellini, G. 2008, MNRAS, 385, 98
\bibitem[2009]{thcek}Tchekhovskoy, A., McKinney,  J. C., \& Narayan, R. 2009, ApJ, 699 ,1789
\bibitem[2007]{tramacere}Tramacere, A., Massaro, F., \& Cavaliere, A. 2007, A\&A, 466, 521
\bibitem[1995]{urry}Urry, C. M., Padovani, P. 1995, PASP, 107, 803
\bibitem[2008]{villata}Villata, M. et al. 2008, A\&A, 481, 79
\bibitem[2009]{vittorini}Vittorini, V. et al. 2009, ApJ, 706, 1433
\end{thebibliography}
\end{document}